# Theoretical current-voltage characteristics of ferroelectric tunnel junctions

H. Kohlstedt,[1] N. A. Pertsev,[1,2] J. Rodríguez Contreras,[1] and R. Waser[1,3]

[1]*Institut für Festkörperforschung and CNI, Forschungszentrum Jülich, D-52425 Jülich, Germany*

[2]*A. F. Ioffe Physico-Technical Institute, Russian Academy of Sciences, 194021 St. Petersburg, Russia*

[3]*Institut für Werkstoffe der Elektrotechnik, RWTH Aachen University of Technology, D-52056 Aachen, Germany*

We present the concept of ferroelectric tunnel junctions (FTJs). These junctions consist of two metal electrodes separated by a nanometer-thick ferroelectric barrier. The current-voltage characteristics of FTJs are analyzed under the assumption that the direct electron tunneling represents the dominant conduction mechanism. First, the influence of converse piezoelectric effect inherent in ferroelectric materials on the tunnel current is described. The calculations show that the lattice strains of piezoelectric origin modify the current-voltage relationship owing to strain-induced changes of the barrier thickness, electron effective mass, and position of the conduction-band edge. Remarkably, the conductance minimum becomes shifted from zero voltage due to the piezoelectric effect, and a strain-related resistive switching takes place after the polarization reversal in a ferroelectric barrier. Second, we analyze the influence of the internal electric field arising due to imperfect screening of polarization charges by electrons in metal electrodes. It is shown that, for asymmetric FTJs, this depolarizing-field effect also leads to a considerable change of the barrier resistance after the polarization reversal. However, the symmetry of the resulting current-voltage loop is different from that characteristic of the strain-related resistive switching. The crossover from one to another type of the hysteretic curve, which accompanies the increase of FTJ asymmetry, is described taking into account both the strain and depolarizing-field effects. It is noted that asymmetric FTJs with dissimilar top and bottom electrodes are preferable for the non-volatile memory applications because of a larger resistance on/off ratio.

## I. INTRODUCTION

The tunnel effect played a significant role during the development of quantum mechanics in the thirties of the last century because it provided a proof for the wave–particle dualism.[1-3] Electron tunneling is described with the aid of electron wave functions determined by the Schrödinger equation. A typical tunnel junction consists of two metal layers separated by a thin insulator (tunnel barrier). Although it is classically forbidden, an electron can traverse a potential barrier with the height exceeding the electron energy. However, the tunneling probability becomes significant only for ultrathin (nanometer-thick) barriers. Excellent textbooks have been published on the subject of quantum mechanical electron tunneling.[4-6]

At present, various types of tunnel junctions are studied from the fundamental point of view and used in microelectronics. Famous examples are the superconducting $Nb/Al-Al_2O_3/Nb$ and magnetic $CoFe/Al_2O_3/CoFe$ tunnel junctions for high-frequency digital electronics[7] and non-volatile memory applications,[8] respectively. Of particular interest for this work is the recent research on epitaxial magnetic oxide tunnel junctions, which are made of $(La_{0.67}Sr_{0.33})CoO_3$ or $(La_{0.67}Ca_{0.33})CoO_3$ electrodes and $SrTiO_3$ or $NdGaO_3$ tunnel barriers and grown on $SrTiO_3$ or $NdGaO_3$ substrates.[9-11] The results obtained for these oxide magnetic tunnel junctions stimulate the development of new types of all-oxide junctions.

The aforementioned metallic and oxide tunnel junctions have two features in common. First, their specific properties are associated with a cooperative phenomenon (superconductivity or magnetism), which occurs in the electrodes. Second, the barrier material in both junction types belongs to the group of non-polar dielectrics, although the material structure is very different (amorphous $Al_2O_3$ vs. single-crystalline $SrTiO_3$). We shall consider another, very interesting type of tunnel junction, where a ferroelectric is employed as the barrier material.

This device, here termed a *ferroelectric tunnel junction* (FTJ), can be used to study the interplay of ferroelectricity and electron tunneling. It may be noted that the discovery of ferroelectricity goes back to 1920,[12] i.e., approximately to the period of time when the principles of quantum mechanical electron tunneling[1] have been formulated.

The idea and very preliminary study of a FTJ (called a polar switch at that time) was presented already in 1971 by Esaki.[13] However, the realization of this idea is a task with many obstacles, because it requires the fabrication of ultrathin films retaining pronounced ferroelectric properties at a thickness of only a few unit cells. Although there are several publications on the electron tunneling in ferroelectrics,[13-18] the experimental studies of the tunneling *across* ferroelectric barriers just started.[19-21] Moreover, we are not aware of any theoretical investigations of the direct electron tunneling through an ultrathin ferroelectric barrier.

Since ferroelectricity is a collective phenomenon like superconductivity and magnetism, thin films are expected to be ferroelectric only above some minimum film thickness. The critical thickness for ferroelectricity has been discussed since early seventies of the last century[22,23] and for a long time was believed to be relatively large (~10-100 nm, see the data collected in Ref. 19). However, the recent work of Tybell et al.[24] demonstrated the presence of a stable polarization in the 4-nm-thick epitaxial film of a perovskite ferroelectric $Pb(Zr_{0.2}Ti_{0.8})O_3$. Experimental evidence has been obtained for the ferroelectric properties of epitaxial $PbTiO_3$ films with a thickness down to 1.2 nm on the basis of structural investigations.[25] The modern theoretical studies also support the existence of ferroelectricity in ultrathin films.[26-30] Thus, the experimental and theoretical results indicate that FTJs may be realized by using an epitaxial ferroelectric layer as a tunnel barrier.

The concept of a FTJ is illustrated in Fig. 1, which shows a simplified band diagram of the metal-ferroelectric-metal heterostructure together with a sketch of the unit cell of a perovskite ferroelectric crystal. Since ferroelectrics possess several specific physical properties, the current-voltage characteristics of FTJs are expected to be different from those of conventional metal-insulator-metal junctions. In particular, the electric-field-induced polarization reversal in a ferroelectric barrier may have a pronounced effect on the conductance of a FTJ. Indeed, the polarization switching alters the sign of polarization charges existing at a given barrier/electrode interface, changes positions of ions in ferroelectric unit cells, and modifies lattice strains inside the barrier. Motivated by these considerations, we carried out the theoretical analysis of current-voltage (*I-V*) relationships that characterize the direct electron tunneling across FTJs.

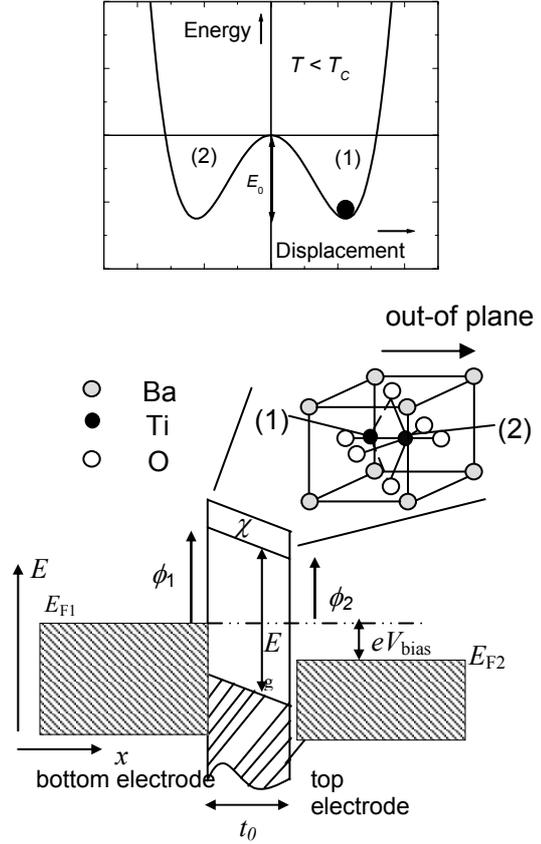

**FIG. 1.** Simplified band diagram of a ferroelectric tunnel junction. $E_F$ is the Fermi energy, $\chi$ is the electron affinity of the ferroelectric, $t$ is the barrier thickness, and $\phi_1$, $\phi_2$ are the barrier heights at the bottom and top electrode, respectively. The inserted sketch shows the structure of a unit cell of $BaTiO_3$, which represents the ferroelectric barrier. Two equilibrium positions of the $Ti^{4+}$ ion are labeled with numbers (1) and (2). The plot shows schematically the variation of ferroelectric energy density with the displacement of $Ti^{4+}$ ions during the polarization reversal.

In this paper, we report our theoretical predictions on the resistive switching and the hysteretic *I-V* curves, which may result from the polarization reversal in FTJs. The barrier material is considered to be a perfect insulator here. For simplicity, we ignore the influence of localized states and structural imperfections, which may be present in a real ferroelectric barrier. Magnitudes of the discussed effects are estimated for oxide ferroelectrics like Pb(Zr$_{0.52}$Ti$_{0.48}$)O$_3$. Ferroelectric polymers represent another opportunity to check the theoretical predictions, because ferroelectricity exists even in a 1-nm-thick polymer film.[31,32] On the microscopic level, ferroelectric properties of polymers are associated with the presence of permanent dipoles so that the situation here is very different from the case of oxide ferroelectrics. The macroscopic effects discussed in this work, however, must exist in all types of FTJs.

Section II is devoted to the theoretical description of the direct quantum mechanical electron tunneling through an insulating barrier possessing piezoelectric properties inherent in ferroelectrics. The calculations show that the voltage-dependent lattice strain, which is caused by the converse piezoelectric effect, modifies the *I-V* relationship owing to strain-induced changes of the barrier thickness, electron effective mass, and position of the conduction-band edge. The results of these calculations are used in Sec. IIIA to predict the strain-related effect of the polarization switching on the *I-V* curves of symmetric FTJs (i.e., junctions with identical top and bottom electrodes). Possible influence of an internal electric field, which may be present in a symmetric FTJ due to the polarization charges existing at the surfaces of a ferroelectric barrier, is discussed in Sec. IIIB. A microscopic interface effect, which is associated with the displacements of ions in ferroelectric unit cells during the polarization switching, is also considered (Sec. IIIC). In Section IV, we analyze the properties of asymmetric FTJs, which involve dissimilar top and bottom electrodes. It is shown that the depolarizing-field effect in asymmetric tunnel junctions may lead to a qualitative change of the hysteretic *I-V* curve (Sec. IVA). The crossover from the strain-dominated resistive switching in symmetric and weakly asymmetric junctions to the depolarizing-field-dominated switching in strongly asymmetric FTJs is finally described (Sec. IVB).

## II. DIRECT ELECTRON TUNNELING THROUGH PIEZOELECTIC BARRIERS

We shall analyze first the process of electron tunneling through an insulating barrier possessing piezoelectric properties. The insulator is supposed to be sandwiched between two identical normal metal electrodes so that at zero voltage the barrier has a rectangular shape. (In this section, we assume that the depolarizing field in a short-circuited FTJ vanishes due to the perfect screening of polarization charges). If such metal-insulator-metal junction is grown epitaxially on a much thicker substrate, the in-plane dimensions of the barrier are totally controlled by the substrate. Accordingly, the application of a potential difference *V* between the electrodes cannot induce any additional in-plane lattice strains in the insulator ($\Delta S_1 = \Delta S_2 = \Delta S_6 = 0$). (We use the Voigt matrix notation and the rectangular reference frame with the $x_3$ axis orthogonal to the substrate surface.) At the same time, the out-of-plane strains $S_3$, $S_4$, and $S_5$ can vary due to the converse piezoelectric effect. For simplicity, we shall assume that the electric field changes the initial barrier thickness $t_0$ but does not create any tilt of the crystal lattice ($\Delta S_3 \neq 0$, $\Delta S_4 = \Delta S_5 = 0$). The voltage dependence of the lattice strain $\Delta S_3$ and the barrier thickness $t$ can be written as $\Delta S_3 = d_{33}^* V/t_0$ and $t = t_0 + d_{33}^* V$, where $d_{33}^*$ is the effective longitudinal piezoelectric coefficient of a clamped epitaxial layer. (It can be calculated as $d_{33}^* = d_{33} - 2 d_{31} s_{13} / (s_{11} + s_{12})$ from the piezoelectric coefficients $d_{in}$ and elastic compliances $s_{mn}$ of the barrier material.[33]) Evidently, the dependence $\Delta S_3(V)$ must modify the current-voltage characteristic of a tunnel junction.

To calculate the current through a piezoelectric barrier, we shall use the Wenzel-Kramer-Brillouin (WKB) approximation and the one-band model. For the tunneling probability, this yields[4-6]

$$D = \exp\left[-2\int_0^t (-K_3^2)^{1/2} dx_3\right], \quad (1)$$

where $K_3$ is the electron wave number normal to the barrier plane. In the one-band model, the total energy $E$ near the bottom $E_c$ of the barrier conduction band (CB) can be written as

$$E = E_c + \frac{\hbar^2 K_1^2}{2m_1^*} + \frac{\hbar^2 K_2^2}{2m_2^*} + \frac{\hbar^2 K_3^2}{2m_3^*}. \quad (2)$$

where $K_1$ and $K_2$ are the wave numbers for in-plane crystallographic directions, $m_i^*$ ($i = 1,2,3$) are the electron effective masses of the insulator, and $\hbar$ is the Planck constant. Equation (2) gives $-K_3^2 = (2m_3^*/\hbar^2)(E_c - E_3)$, where $E_3$ is the total electron energy in the direction perpendicular to the barrier. The CB edge $E_c$ involved in this relation depends on the strain state of the crystal lattice.[34] For our purposes, it is sufficient to write the strain dependence of $E_c$ as $E_c = E_c^0 + \kappa_3 \Delta S_3$, where $E_c^0$ is the minimum of the conduction band in a constrained barrier at $V = 0$, and $\kappa_3$ is the relevant deformation potential of the conduction band. The electron effective mass also changes in the presence of lattice strains[35] so that we have $m_3^* = m_3^{*0}(1 + \mu_{33}\Delta S_3)$, where $m_3^{*0} = m_3^*(\Delta S_3 = 0)$ and $\mu_{33} = \partial \ln m_3^*/\partial S_3$. The described strain effects result in the voltage dependences of the CB edge $E_c = E_c^0 + \kappa_3 d_{33}^* V/t_0$ and the electron effective mass $m_3^* = m_3^{*0}(1 + \mu_{33} d_{33}^* V/t_0)$. Taking into account the energy change ($-eVx_3/t$) during the motion of an electron charge $e$ in an electric field $V/t$, and allowing for the voltage dependences of $E_c$, $m_3^*$, and the barrier thickness $t = t_0 + d_{33}^* V$, from Eq. (1) we find the tunneling probability as

$$D(E_3) = \exp\left\{-\frac{4(2m_3^{*0})^{1/2}}{3\hbar e}\frac{t_0}{V}\left(1 + \frac{d_{33}^* V}{t_0}\right)\left(1 + \mu_{33}\frac{d_{33}^* V}{t_0}\right)^{1/2}\right.$$
$$\left.\times\left[\left(E_c^0 + \kappa_3\frac{d_{33}^* V}{t_0} - E_3\right)^{3/2} - \left(E_c^0 - eV + \kappa_3\frac{d_{33}^* V}{t_0} - E_3\right)^{3/2}\right]\right\} \quad (3)$$

It can be seen that the voltage dependence of $D$ differs considerably from the case of a nonpiezoelectric barrier.[4-6] For small voltages, Eq. (3) may be expanded in powers of the voltage with the account of the first three terms only.

For simplicity, we shall calculate the current through a piezoelectric barrier at $T = 0$ K using the approximation developed by Simmons.[36] The integral expression for the current density $J = I/A$ reads

$$J = \frac{4\pi me}{h^3}\left\{eV\int_0^{E_F - eV} D(E_3)dE_3 + \int_{E_F - eV}^{E_F}(E_F - E_3)D(E_3)dE_3\right\}, \quad (4)$$

where $E_F$ is the Fermi level of the electrodes, and $m$ is the free electron mass. The integration in Eq. (4) has been performed for the case of small voltages $V$ by substituting the series expansion of $D$ obtained from Eq. (3). Using the program MATHEMATICA®, we determined the linear, quadratic, and cubic terms in the series expansion $J(V) = C_1 V + C_2 V^2 + C_3 V^3 + ...$ The coefficient $C_1$ of the linear term, which is unaffected by the piezoelectric effect, equals

$$C_1 = \frac{me^2}{4\pi\hbar^2 m_3^{*0} t_0^2}\left[4\pi t_0 (2m_3^{*0})^{1/2}\phi_0^{1/2} + h\right]$$
$$\times \exp\left[-\frac{4\pi}{h}t_0(2m_3^{*0})^{1/2}\phi_0^{1/2}\right], \quad (5)$$

where $\phi_0 = E_c^0 - E_F$ is the barrier height at $V = 0$. This expression is similar to that given by Brinkman et al.[37] The quadratic term in the current-voltage dependence, which is negligible for a symmetric barrier,[36,37] becomes nonzero in the presence of piezoelectricity. The corresponding coefficient is given by the formula

$$C_2 = -\frac{4\pi m e^2}{h^3} d_{33}^* \left\{ \left[ \frac{\phi_0}{t_0} + \frac{h\phi_0^{1/2}}{2\pi(2m_3^{*0})^{1/2} t_0^2} + \frac{h^2}{16\pi^2 m_3^{*0} t_0^3} \right] \right.$$
$$\left. \times (2 + \mu_{33}) + \frac{\kappa_3}{t_0} \right\} \exp\left[ -\frac{4\pi}{h} t_0 (2m_3^{*0})^{1/2} \phi_0^{1/2} \right]. \quad (6)$$

It can be seen that $C_2$ is proportional to the piezoelectric constant $d_{33}^*$. The last term in curly brackets describes the influence of the band-edge shift. Other terms determine the combined effect of the strain-induced changes of barrier thickness and effective mass. The calculation of the coefficient $C_3$ leads to the expression

$$C_3 = \frac{me^2}{4\pi h^4 t_0^4 m_3^{*0} \phi_0^{1/2}} \left\{ \frac{4}{3}\pi^3 e^2 t_0^5 (2m_3^{*0})^{3/2} + h d_{33}^{*2}(3 + 2\mu_{33} + \mu_{33}^2)\phi_0^{1/2} \left[ h^2 + 4\pi h t_0 (2m_3^{*0})^{1/2}\phi_0^{1/2} + 16\pi^2 t_0^2 m_3^{*0} \phi_0 \right] \right.$$
$$\left. + 8\pi^3 d_{33}^{*2} t_0^3 (2m_3^{*0})^{3/2} [(2 + \mu_{33})\phi_0 + \kappa_3]^2 \right\} \exp\left[ -\frac{4\pi}{h} t_0 (2m_3^{*0})^{1/2} \phi_0^{1/2} \right] + \frac{8\pi^3 m e^4 t_0^2 m_3^{*0}}{3h^5} \text{Ei}\left[ -\frac{4\pi}{h} t_0 (2m_3^{*0})^{1/2} \phi_0^{1/2} \right], \quad (7)$$

where Ei($z$) is the exponential integral function, $\text{Ei}(z) = -\int_{-z}^{\infty} [\exp(-x)/x] dx$ for $z > 0$. Equation (7) shows that the piezoelectric effect always increases the cubic term, irrespective of the sign of $d_{33}^*$.

The most remarkable manifestation of the barrier piezoelectric properties is an asymmetry of the current-voltage characteristic with respect to $V = 0$. Since the quadratic term differs from zero, the conductance $G(V) = dJ/dV$ becomes minimal at the voltage $V_{\min} = -C_2/(3C_3)$. Therefore, the conductance minimum is shifted from zero voltage, in contrast to a nonpiezoelectric rectangular barrier.[37] To evaluate the significance of the predicted effect, we performed numerical calculations of the conductance $G(V)$. The piezoelectric coefficient $d_{33}^*$ is expected to be large in films of Pb(Zr$_{1-x}$Ti$_x$)O$_3$ (PZT) solid solutions with compositions near the bulk morphotropic boundary ($x \approx 0.5$).[38] Taking into account the expected reduction of piezoelectric response in ultrathin films, we assumed $d_{33}^* = 50$ pm/V. Since the CB deformation potential is unknown for PZT, we used the theoretical value $\kappa_3 = -4.6$ eV obtained in Ref. 39 for wurtzite GaN, which also possesses piezoelectric properties. For the strain sensitivity $\mu_{33}$ of the effective mass, we assumed $\mu_{33} = 10$ on the basis of the calculated strain-induced change of the effective mass in wurtzite Ga$_{0.7}$Al$_{0.3}$N.[40]

Since $\mu_{33}$ is positive, the voltage-induced variations of the effective mass $m_3^*$ and barrier thickness $t$ have a similar effect on the conductance [see Eq. (3)]. The band-edge shift, however, influences $G(V)$ in an opposite way at $\kappa_3 < 0$, as demonstrated by Eq. (6). The sign of the total effect depends on the magnitudes of involved material parameters and on the barrier height $\phi$ and thickness $t_0$. According to our numerical calculations, at $\phi_0 = 0.5$ eV, $t_0 = 2$ nm, $m_3^{*0} = 0.2\,m$ (as calculated for GaN in Ref. 39), and the above values of $\mu_{33}$ and $\kappa_3$, the combined effect of effective-mass and barrier-thickness changes prevails over that of the band-edge shift. As a result, the conductance decreases when the voltage induces tensile out-of-plane strain $\Delta S_3$ in the barrier and increases at the appearance of a compressive strain $\Delta S_3(V)$.

The calculated voltage dependence of the current density $J$ through a piezoelectric film and that of the barrier conductance $G$ are shown in Fig. 2. Since the applied electric field may induce either tensile or compressive strain $\Delta S_3$ in the piezoelectric layer, two distinct curves $J(V)$ can be observed, depending on the face of the barrier to which positive voltage is applied. Accordingly, the conductance minimum may be shifted either to a positive or to a negative voltage $V_{\min}$ [Fig. 2(b)]. The numerical calculation gives $|V_{\min}| \approx 50$ mV, which is comparable with the offsets caused by the use of two different electrodes in conventional junctions.[37]

## III. RESISTIVE SWITCHING IN SYMMETRIC FERROELECTRIC JUNCTIONS

### A. Strain effect

Ferroelectric materials are distinguished from other piezoelectrics by the presence of a spontaneous polarization that can be switched by an applied electric field.[41] On this stage of our theoretical analysis, it is sufficient to employ the simplest model of the switching process, which assumes that the polarization reversal occurs simultaneously in the whole single-domain film at a critical electric field $\mathscr{E}_c$. This coercive field $\mathscr{E}_c$ is generally a thickness-dependent characteristic of the film,[42-44] which determines the *coercive voltage* $V_c = \mathscr{E}_c t_0$. Before the polarization switching ($V < V_c$), the piezoelectric coefficient $d_{33}^*$ is negative since the polarization **P** is directed against the applied field $\mathscr{E}$. After the polarization reversal ($V \geq V_c$), the coefficient $d_{33}^*$ changes its sign from negative to positive, which is accompanied by a step-like increase in the film thickness $t$ and the lattice strain $S_3$ (by the amounts $\delta t = 2|d_{33}^*|V_c$ and $\delta S_3 = 2|d_{33}^*|V_c/t_0$, respectively). As a result, the strain-voltage relationship $\Delta S_3(V)$ of a ferroelectric film demonstrates the hysteresis shown schematically in Fig. 3 and known as "butterfly" curve.[45]

Owing to the strain changes accompanying the polarization reversal, the current-voltage characteristic of a FTJ must also exhibit a hysteretic behavior and resistive switching, as shown in Fig. 2(a). At both negative and positive coercive voltages $\pm V_c$, the barrier conductance $G$ experiences a step-like drop [Fig. 2(b)]. In our approximation, the jump of conductance is caused solely by the change of sign of the coefficient $C_2$, because $C_1$ and $C_3$ are not sensitive to the sign of $d_{33}^*$ [see Eqs. (5)-(7)].

It should be emphasized that the predicted *strain-related resistive switching* has the following specific feature. Evidently, the high-resistance state of a FTJ at all voltages corresponds to the polarization orientation parallel to the applied electric field $\mathscr{E}$, whereas the low-resistance state is associated with the polarization oriented against $\mathscr{E}$. Therefore, at zero voltage an "inversion" of the junction resistance state takes place, although nothing happens with the FTJ itself. Indeed, the high-resistance state becomes the low-resistance one or vise versa [see Fig. 2(b)], because the direction of applied field reverses at $V = 0$. A similar diode-like behavior of a FTJ was suggested in Ref. 14.

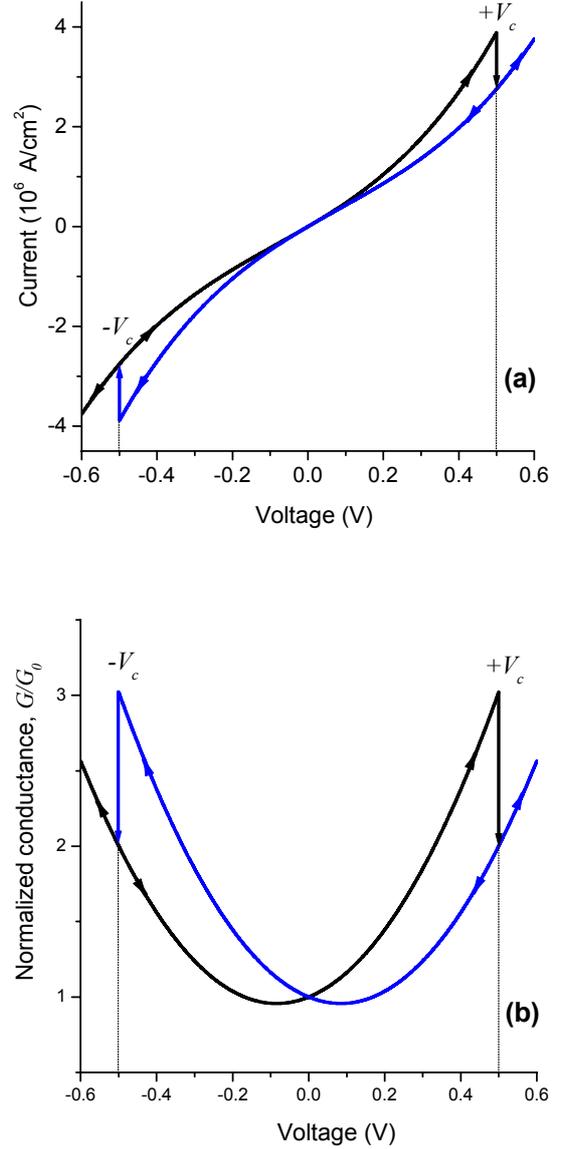

**FIG. 2.** Influence of converse piezoelectric effect on the current-voltage (a) and conductance-voltage (b) characteristics of tunnel junctions. Theoretical calculations of the current density $J$ and conductance $G$ per unit area were performed using the following values of the junction parameters: $\phi_0 = 0.5$ eV, $t_0 = 2$ nm, $m_3^{*0} = 0.2\ m$, $d_{33}^* = 50$ pm/V, $\kappa_3 = -4.6$ eV, and $\mu = 10$. The conductance is normalized by its value at zero voltage, $G_0 = G(V = 0)$. The resistive switching at voltages $\pm V_c$ and the resulting hysteretic behavior correspond to the case of a ferroelectric tunnel junction, where the polarization reversal takes place in the barrier at the coercive voltage $V_c$.

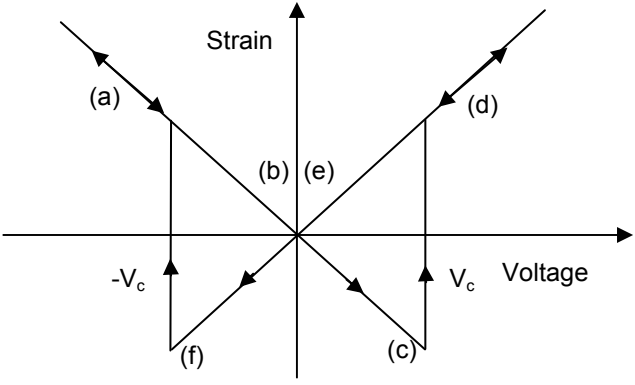

**FIG. 3.** Dependence of the out-of-plane lattice strain $\Delta S_3$ in an epitaxial ferroelectric film on the applied voltage $V$ (simplified representation). The complete cycle is shown by letters from (a) to (f).

### B. Depolarizing-field effect

In addition to the resistive switching caused by the strain effect associated with piezoelectricity, other mechanisms may be proposed for the influence of polarization reversal on the electron transport across ferroelectric barriers. The most evident mechanism is related to the possible presence of a *depolarizing electric field* in a FTJ. This field is created by the polarization charges $\rho = -\,\text{div}\,\mathbf{P}$ existing at the film surfaces. Even in the case when these surfaces are covered by metal electrodes, the depolarizing field may differ from zero due to the finite electronic screening length in metals.[29,46] Evidently, this internal electric field modifies the potential barrier in the FTJ and so may change the tunneling current. (It should be noted that the electric fields induced by polarization charges may strongly influence the charge transport in ferroelectric/semiconductor heterostructures.[47,48])

The recent first-principles calculations[29] demonstrated that the internal electric field in a short-circuited symmetric metal-ferroelectric-metal heterostructure is roughly constant within the ferroelectric layer. The electric field also exists inside subsurface layers of the electrodes, where it has the opposite sign [see Fig. 4(a)]. Accordingly, the distribution of an electrostatic potential across a short-circuited symmetric FTJ can be approximated by the zigzag profile shown in Fig. 4(b). The potential profile of this type changes neither the mean barrier height nor the average slope of the barrier. In symmetric FTJs, therefore, the depolarizing field cannot induce significant resistive switching. Moreover, at zero voltage an "inversion" of the potential profile seen by the tunneling electrons takes place so that the depolarizing-field effect could only produce the same symmetry of hysteretic $I$-$V$ curves as the strain effect.

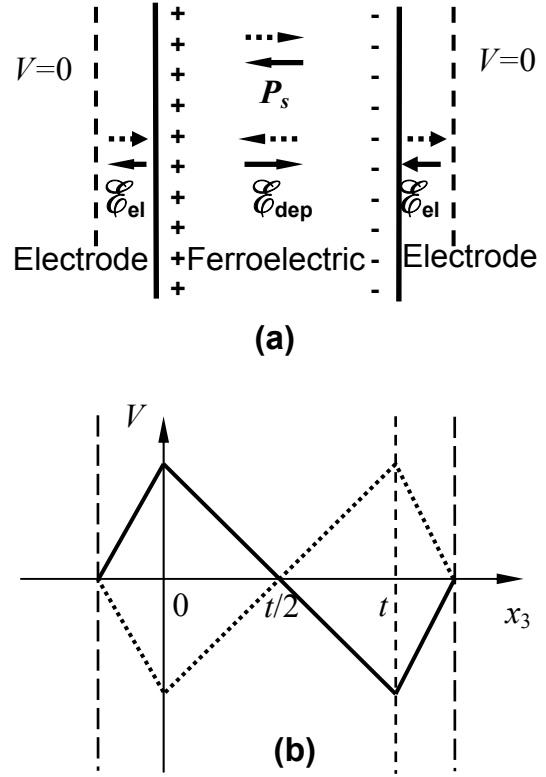

**FIG. 4.** Internal electric fields $\mathscr{E}_{\text{dep}}$ and $\mathscr{E}_{\text{el}}$ in a short-circuited symmetric FTJ (a) and the model distribution of an electrostatic potential $V$ across this junction (b). Two possible polarization states and the corresponding potential profiles are shown by solid and dotted arrows and lines. The penetration of the electric field into the electrodes is determined by the screening length of electrode material.

### C. Microscopic interface effect

In addition, a *microscopic interface effect* may be predicted, which is due to the displacements of ions in ferroelectric unit cells during the polarization switching. To describe this effect qualitatively, we consider a perovskite ferroelectric like $BaTiO_3$ or $PbTiO_3$. In the tetragonal ground state, the $Ti^{4+}$ ion is shifted with respect to the center of the unit cell, as shown schematically in Fig. 1. After the polarization reversal, the direction of this shift changes to the opposite one so that the distance between subsurface $Ti^{4+}$ ions and the electrode increases or decreases by some amount. These displacements of $Ti^{4+}$ ions modify the microscopic structure of an interfacial region, which may lead to a change of the barrier height at the given electrode. The two possible situations are shown schematically in Fig. 5 for a FTJ involving $SrRuO_3$ electrodes with SrO terminations at both interfaces and $BaTiO_3$ as a ferroelectric barrier.

Owing to the structural asymmetry of a poled ferroelectric film, the barrier heights $\phi_1$ and $\phi_2$ at the bottom and top electrodes in a FTJ must differ from each other. For the qualitative description of this interface effect on the tunnel current, we may assume that the barrier has a trapezoidal shape $\phi(x_3) = \overline{\phi} + \Delta\phi\,(x_3 - \tfrac{1}{2} t_0)/t_0$ at $V = 0$, where $\overline{\phi} = (\phi_1 + \phi_2)/2$ is the mean barrier height, and $\Delta\phi = (\phi_2 - \phi_1)$ is the barrier asymmetry. The polarization reversal transforms the barrier profile into $\phi(x_3) = \overline{\phi} - \Delta\phi\,(x_3 - \tfrac{1}{2} t_0)/t_0$ so that a "reflection" of $\phi(x_3)$ with respect to the barrier center $x_3 = \tfrac{1}{2} t_0$ takes place.

When modeling the interface-related effect of polarization switching on the electron tunneling, we may neglect the influence of applied voltage on the film thickness and the barrier heights $\phi_1$ and $\phi_2$ at the electrodes. Then the problem becomes equivalent to the calculation of the tunnel current through a trapezoidal potential barrier, which was performed by Brinkman *et al.*[37] In the low-voltage range, the current density may be approximated as $J(V) = C_1 V + C_2 V^2 + C_3 V^3$, where the coefficients $C_1$ and $C_3$ are independent of the barrier asymmetry $\Delta\phi$, whereas $C_2$ is directly proportional to $\Delta\phi$.[37] Since the polarization reversal simply changes the sign of $\Delta\phi$ in our case, the interface-related effect on the electron tunneling through a ferroelectric barrier appears to be qualitatively similar to the strain-related one. If this effect is strong enough, it will also create a step-like change of the barrier conductance $G$ at the coercive voltage and produce the symmetry of $J(V)$ and $G(V)$ curves shown in Figs. 2(a) and 2(b), respectively.

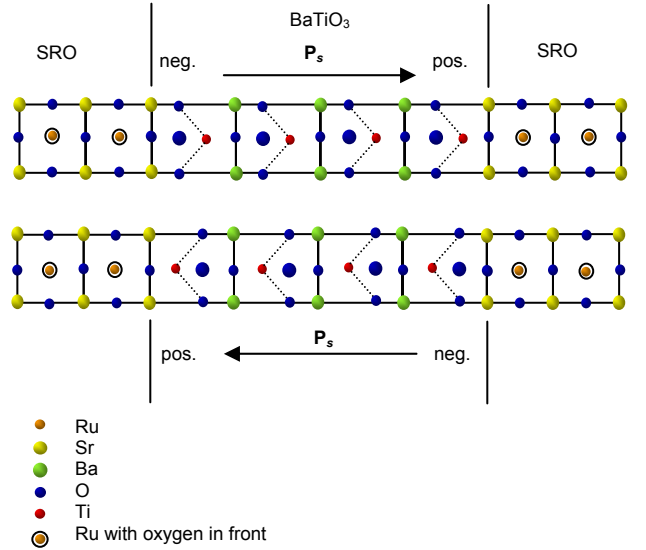

**FIG. 5.** Schematic representation of the atomic structure of a $SrRuO_3/BaTiO_3/SrRuO_3$ trilayer. Two possible configurations are shown, which correspond to opposite directions of the spontaneous polarization $\mathbf{P}_s$. $SrRuO_3$ electrodes are assumed to have SrO terminations at both interfaces. Note different signs of the polarization charges and different positions of atoms at two interfaces.

It should be noted that the predicted microscopic interface effect on the direct electron tunneling may manifest itself in the case of piezoelectric (but not ferroelectric) barriers as well. Although the resistive switching does not occur in the absence of field-induced polarization reversal, two different *I-V* characteristics can be observed by applying the positive voltage to the opposite faces of piezoelectric barrier (i.e., to the top or bottom electrode). The difference in transport properties is again caused by different barrier heights at the top and bottom electrodes, which result from different atomic structures of two subsurface layers of a piezoelectric crystal. The situation here has some analogy with the case of Pt/GaN Schottky diodes, where the *I-V* characteristics were found to be different for GaN layers with Ga-terminated and N-terminated surfaces.[49] However, the difference in the barrier heights of these two Pt/GaN Schottky contacts is believed to be caused by the opposite sign of the polarization charges existing at Ga- and N-faces.[49,50]

## IV. RESISTIVE SWITCHING IN ASYMMETRIC JUNCTIONS

In this section, we consider the case of asymmetric junctions, which involve dissimilar top and bottom electrodes like Pt/Pb(Zr$_{0.52}$Ti$_{0.48}$)O$_3$/SrRuO$_3$ junctions studied in Ref. 21. It is of fundamental interest and of practical importance to analyze how such asymmetry influences the current-voltage characteristics of FTJs.

When two different metals are used to produce the top and bottom electrodes in a FTJ, two consequences are evident. First, the difference in the work functions of electrodes leads to the appearance of an internal electric field inside the film, which transforms the rectangular barrier into a trapezoidal one.[51] Second, since the abilities of the top and bottom electrodes to screen the depolarizing field $\mathscr{E}_{dep}$ are different, the distribution of an electrostatic potential in the FTJ becomes more asymmetric than the distribution shown in Fig. 4. Evidently, the "work-function" effect alone cannot change the symmetry of the hysteretic *I-V* curves. Therefore, we focus on the role of the depolarizing field in an asymmetric FTJ.

### A. Depolarizing-field effect in asymmetric junctions

The simplest electrostatic model of an asymmetric FTJ may be constructed under the assumption that the screening charge in one of the electrodes is located on a plane shifted from the ferroelectric surface by a finite distance $\Delta t$, whereas the other electrode provides perfect screening of polarization charges ($\Delta t' = 0$). In the space $\Delta t$ between two charges, the electric field $\mathscr{E}_{int}$ is opposite to the depolarizing field $\mathscr{E}_{dep}$ (see Fig. 6), and the material has a finite permittivity. The introduced interfacial layer either models a poor screening of $\mathscr{E}_{dep}$ by one of the electrodes[52] or represents a real nonferroelectric layer, which is believed to form in some ferroelectric/metal heterostructures (e.g., in capacitors involving the Pt electrode).[43,44,53-55] The distribution $V(x_3)$ of the electrostatic potential in a short-circuited FTJ containing an interfacial layer is shown in Fig. 6. (Here we put aside the work-function effect, assumed the barrier to be an ideal insulator, and neglected the contribution to $\mathscr{E}_{dep}$ created by the second interface.) It can be seen that, in contrast to the case of a symmetric FTJ (Fig. 4), the mean value of the electrostatic potential in the barrier is not equal to zero, and, moreover, it changes sign after the polarization reversal.

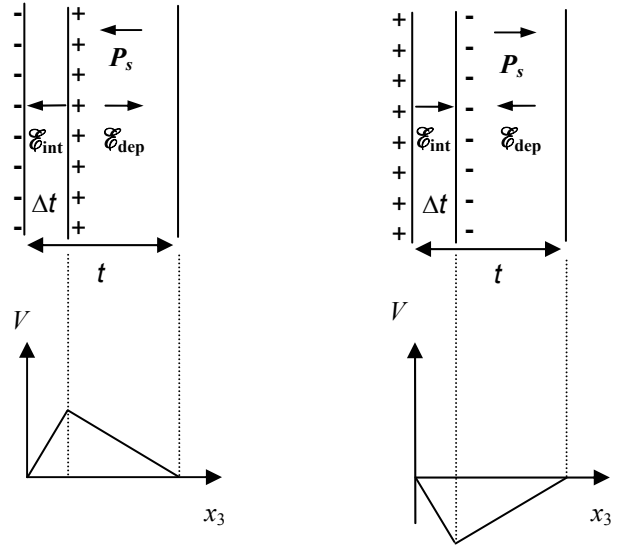

**FIG. 6.** Internal electric fields $\mathscr{E}_{dep}$ and $\mathscr{E}_{int}$ in a short-circuited asymmetric FTJ (a) and the model distribution of an electrostatic potential across this junction (b). The junction contains an effective (or real) interfacial layer of thickness $\Delta t$ at one of the film/electrode interfaces. The model corresponds to a junction involving two different electrodes (e.g., Pt and SRO). The screening abilities of electrodes are assumed to be very different, but their work functions are taken to be the same. Two possible potential profiles are shown, which correspond to opposite orientations of the spontaneous polarization **P**$_s$ in a ferroelectric film.

A depolarizing-field modification of the potential barrier for electron tunneling can be described using the following formula for the barrier profile $\phi(x_3)$ at $V = 0$:

$$\phi(x_3) = \begin{cases} \phi_0 - e\mathcal{E}_{\text{dep}}(t-\Delta t)x_3/\Delta t & \text{at } 0 \leq x_3 \leq \Delta t, \\ \phi_0 + e\mathcal{E}_{\text{dep}}(x_3 - t) & \text{at } \Delta t \leq x_3 \leq t, \end{cases} \quad (8)$$

where $\phi_0$ is the barrier height in the absence of the depolarizing field $\mathcal{E}_{\text{dep}}$ in the ferroelectric material, $t$ is the total barrier thickness, and $\Delta t$ is the thickness of an interfacial layer. The magnitude of $\mathcal{E}_{\text{dep}}$ depends on the out-of-plane polarization $P_3$ in the ferroelectric film. From the continuity condition for the electric displacement and the condition of zero average field in a short-circuited FTJ we obtain $\mathcal{E}_{\text{dep}} \cong -P_3/[c_i(t - \Delta t)]$, where $c_i$ is the capacitance density associated with the interfacial layer (or the dielectric/metal interface itself[52]). The equilibrium polarization $P_3$ in a strained epitaxial film without in-plane polarization components ($P_1 = P_2 = 0$) can be calculated from the equation $\partial \tilde{G}/\partial P_3 = \mathcal{E}_{\text{dep}}$, where $\tilde{G}$ is the modified thermodynamic potential introduced in Ref. 56. The calculation shows that the depolarizing-field effect is equivalent here to a simple shift of the bulk Curie-Weiss temperature $\theta$ by $\Delta\theta \cong -2\varepsilon_0 B/[c_i(t-\Delta t)]$, where $\varepsilon_0$ is the permittivity of the vacuum, and $B$ is the Curie-Weiss constant of the bulk material. Well below the reduced temperature of ferroelectric phase transition, we may use the linear approximation to estimate the equilibrium polarization as $P_3 \cong P_s + \varepsilon_0 \varepsilon_f \mathcal{E}_{\text{dep}}$ ($P_s$ is the spontaneous polarization in the absence of $\mathcal{E}_{\text{dep}}$, and $\varepsilon_f$ is the relative out-of-plane permittivity of a ferroelectric film). This formula yields the relation $\mathcal{E}_{\text{dep}} \cong -P_s/[c_i(t-\Delta t) + \varepsilon_0 \varepsilon_f]$, which enables us to evaluate the depolarizing field.

The integration of $\phi(x_3)$ given by Eq. (8) shows that the mean barrier height $\bar{\phi}$ equals

$$\bar{\phi} = \phi_0 + \frac{1}{2}e(t-\Delta t)\mathcal{E}_{\text{dep}} \cong \phi_0 - \frac{1}{2}\frac{eP_s}{[c_i + \varepsilon_0\varepsilon_f/(t-\Delta t)]}. \quad (9)$$

It can be seen that the polarization reversal changes the mean barrier height by the amount $2\Delta\bar{\phi} \cong e|P_s|(t-\Delta t)/[c_i(t-\Delta t)+\varepsilon_0\varepsilon_f]$. For PZT films, assuming $t = 2$ nm, $\Delta t \ll t$, $P_s = 0.5$ C/m$^2$, and $c_i < 0.5$ F/m$^2$, we obtain $\Delta\bar{\phi} > 0.1$ eV at $\varepsilon_f < 300$. Since a change of the mean barrier height $\bar{\phi}$ strongly affects the tunnel current,[37] we see that the depolarizing-field effect may influence the FTJ conductance significantly. The increase or decrease of the barrier resistance after the polarization reversal depends on the position of the depolarizing-field source in a junction (at the positively biased electrode or at the negatively biased one). It should be noted that the depolarizing-field effect is not expected to induce significant diode-like behavior near $V = 0$.

To evaluate the effect of depolarizing field on the barrier conductance $G$ at zero voltage, we shall use the approximation of an average barrier introduced by Simmons.[36] Replacing the actual barrier profile $\phi(x_3)$ by the rectangular barrier with a height $\bar{\phi}$, we can find $G(V = 0) = C_1$ from Eq. (5) by substituting $\bar{\phi}$ for $\phi_0$. Hence for the ratio of the conductances $G_L$ and $G_H$, which characterize the low- and high-resistance states, we obtain

$$\frac{G_L(V=0)}{G_H(V=0)} \cong \left[\frac{4\pi t_0 (2m_3^{*0})^{1/2}(\phi_0 - \Delta\bar{\phi})^{1/2} + h}{4\pi t_0 (2m_3^{*0})^{1/2}(\phi_0 + \Delta\bar{\phi})^{1/2} + h}\right]$$
$$\times \frac{\exp\left[-\frac{4\pi}{h}t_0(2m_3^{*0})^{1/2}(\phi_0 - \Delta\bar{\phi})^{1/2}\right]}{\exp\left[-\frac{4\pi}{h}t_0(2m_3^{*0})^{1/2}(\phi_0 + \Delta\bar{\phi})^{1/2}\right]}. \quad (10)$$

The detailed calculations based on Eqs. (1) and (4) support the validity of this relation at $\Delta\bar{\phi} \ll \phi_0$. With $\phi_0 = 0.5$ eV, $\Delta\bar{\phi} = 0.1$ eV, $t_0 = 2$ nm, and $m_3^{*0} = 0.2\,m$ (see Sec. II), Eq. (10) gives $G_L/G_H \approx 3$ at $V = 0$. This value of the conductance ratio is large enough to be detected experimentally.

### B. Crossover between two types of hysteretic current-voltage curves

To calculate the whole current-voltage characteristic of an asymmetric FTJ, one should take into account both the depolarizing-field and strain effects on the electron tunneling. In line with the approximation used in the preceding subsection, we neglect the influence of the depolarizing field on the barrier shape. Accordingly, the two opposite polarization states of the ferroelectric layer are modeled here by the rectangular (at zero voltage) barriers of different heights $\phi_1 = \phi_0 + \Delta\bar\phi$ and $\phi_2 = \phi_0 - \Delta\bar\phi$. In this approximation, the current density $J$ may be calculated using Eqs. (5)-(7) with the parameter $\phi_0$ replaced by $\phi_1$ or $\phi_2$ and a positive or negative piezoelectric constant $d_{33}^*$. Hence the two branches of a hysteretic $I$-$V$ curve can be determined.

By changing the strength $\Delta\bar\phi$ of the depolarizing-field effect at fixed values of the other involved physical parameters, we can find out how the degree of junction asymmetry influences its current-voltage characteristic. Performing the numerical calculations of $J(V)$ at $\Delta\bar\phi$ ranging from 0.02 to 0.1 eV, we obtained a set of characteristics shown in Figs. 7 and 8. It can be seen that the hysteretic $I$-$V$ curve displayed by a "strongly" asymmetric FTJ ($\Delta\bar\phi$ = 0.1 eV) differs drastically from that of the symmetric junction [compare Figs. 8 and 2(a)]. Indeed, the conductance of the symmetric FTJ changes in the same way (drops down) at the positive and negative coercive voltages, whereas the asymmetric junction demonstrates conductance jumps of opposite sign at these voltages. The second distinction is related to the current variations near zero voltage. For symmetric FTJs, two branches of a hysteretic $I$-$V$ loop just touch each other at $V$ = 0, while the crossing of these branches occurs at this voltage in the case of asymmetric junctions.

As the degree of the junction asymmetry increases, the crossover from one to another type of the hysteretic $I$-$V$ curve takes place. In "weakly" asymmetric junctions ($\Delta\bar\phi$ = 0.02 eV) the low-resistance state transforms into the high-resistance one at both switching voltages $\pm V_c$ [see Fig. 7(a)], which is similar to the behavior of symmetric FTJs. However, the current-voltage loop displayed by these asymmetric junctions is distinguished by the *double-crossing* of the branches. The first crossing takes place at $V = 0$, and the second one occurs at a negative voltage of $V^* = -0.26$ V. The increase of the depolarizing-field strength $\Delta\bar{\phi}$ results in a shift of the second crossing point to larger negative voltages. At some value of $\Delta\bar{\phi}$, the "crossing" voltage $V^*$ becomes equal to the negative coercive voltage so that the current jump at $-V_c$ disappears. For our set of the junction parameters, this threshold situation occurs at $\Delta\bar{\phi} \approx 0.03$ eV [see Fig. 7(b)]. At larger degrees of the junction asymmetry, two branches of the hysteretic *I-V* curve cross each other only at zero voltage, and the current jumps at $\pm V_c$ are opposite in sign, as shown in Fig. 7(c).

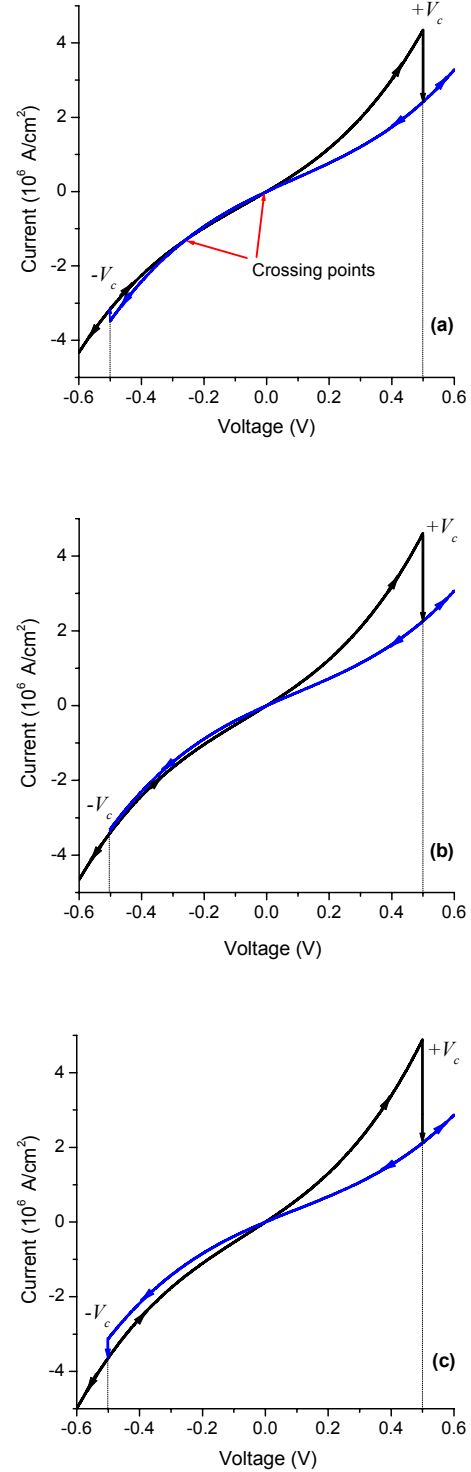

**FIG. 7.** Combined effect of piezoelectric strain and depolarizing field on the $I$-$V$ curves of asymmetric FTJs. The strength $\Delta\bar\phi$ of the depolarizing-field effect is assumed to be 0.02 eV (a), 0.03 eV (b), and 0.04 eV (c). Other junction parameters are taken to be $\phi_0$ = 0.5 eV, $t_0$ = 2 nm, $m_3^{*0}$ = 0.2 $m$, $d_{33}^*$ = 50 pm/V, $\kappa_3$ = –4.6 eV, and $\mu$ = 10. The source of depolarizing field is situated at the biased electrode (another electrode is grounded).

It should be emphasized that magnitudes of the current jumps occurring at the positive and negative switching voltages coincide only in the case of symmetric FTJs with identical top and bottom electrodes [Fig. 2(a)]. Even a small asymmetry results in a significant difference in these magnitudes, as demonstrated by Fig. 7(a). Moreover, the absolute value of a current jump at the positive bias still exceeds considerably the jump at the negative bias even when the junction becomes strongly asymmetric. Indeed, the situation reverses, if the bias voltage is applied to the opposite electrode in the same FTJ (see Fig. 8).

## V. CONCLUDING REMARKS

Our theoretical calculations of the tunnel current across a FTJ strongly support the idea that the polarization reversal in a ferroelectric barrier may result in a pronounced resistive switching. The hysteretic *I-V* curves predicted for symmetric junctions, which involve identical top and bottom electrodes, are distinguished by the absence of the crossing of two branches and by the same conductance jumps at the positive and negative switching voltages. Besides, the tunnel barriers associated with the opposite polarization states in a symmetric junction become identical at zero voltage. Therefore, the ratio of the conductances $G_L$ and $G_H$, which characterize the low- and high-resistance states, goes to unity at $V = 0$.

For asymmetric FTJs, a qualitatively different hysteretic behavior is expected. First, two branches of the

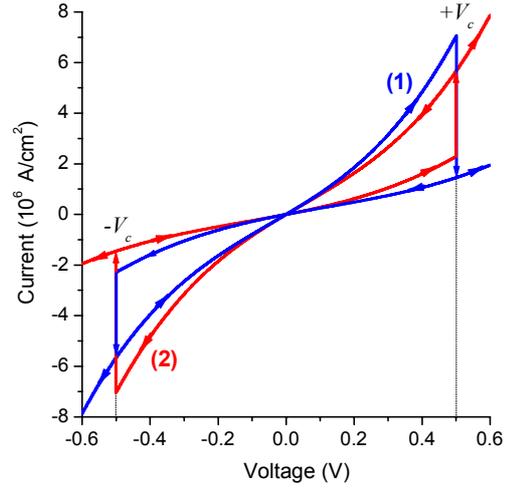

**FIG. 8.** Current-voltage curves of "strongly" asymmetric ferroelectric tunnel junctions ($\Delta\bar{\phi} = 0.1$ eV). Two curves here show the characteristics of FTJs with the source of depolarizing field situated at the biased electrode (1) or at the grounded electrode (2).

*I-V* curve cross at zero voltage so that the ratio $G_L/G_H(V = 0)$ may be several times larger than unity. Second, the current jumps occurring at the positive and negative switching voltages are opposite in sign. These features are caused by the influence of the depolarizing field, which changes the mean barrier height in an asymmetric FTJ. It should be noted that a similar depolarizing-field effect on the conductivity has been proposed earlier for the charge transport inside conductive ferroelectric films containing non-ferroelectric layers.[57]

For memory applications, the asymmetric ferroelectric tunnel junctions seem to be preferable because such junctions can exhibit a larger conductance on/off ratio. This conclusion follows from the fact that here, in contrast to symmetric FTJs, the mean barrier height is different for the two polarization states of a ferroelectric layer even at zero voltage. However, an internal-field-induced shift of the *I-V* curves along the voltage axis must be also taken into account, if the top and bottom electrodes have different work functions. At the same time, even apparently symmetric junctions, where both electrodes are made of the same material, may be asymmetric on the microscopic level. Such asymmetry was revealed recently in some of the $SrRuO_3/BaTiO_3/SrRuO_3$ heterostructures with the aid of the high-resolution transmission electron microscopy.[58] It was found that the bottom $BaTiO_3/SrRuO_3$ interface differs from the upper one by the presence of a Ruddelsen-Popper interfacial layer.[58] Although this asymmetry was engineered by using appropriate deposition conditions, it seems to be a very difficult task to obtain the top and bottom film/electrode interfaces in a FTJ with exactly the same electronic properties. It is likely that slight differences in the physical (e.g., lattice strain) and chemical (termination layer) properties of two interfaces lead to different electronic properties and, therefore, to an asymmetric heterostructure.

## ACKNOWLEDGMENTS

The research described in this publication was supported in part by the HGF-Strategiefonds "Piccolo" and by the Volkswagen-Stiftung Project "Nano-sized ferroelectric hybrids" under Contract No. I/77737. The financial support of the Deutsche Forschungsgemeinschaft is also gratefully acknowledged. J. Rodriguez Contreras likes to thank the German academic exchange service (DAAD) for supporting his stay at the Penn State University in the group of Darrel Schlom.

We acknowledge Jürgen Schubert, Andreas Gerber, Darrell Schlom, Paul Müller, René Meyer, Philippe Ghosez, Christoph Buchal, Ulrich Poppe, Paul Meuffels, Marin Alexe, Steven Ducharme, Michael Indlekofer, Peter Dowben, Ravi Droopad, Karl-Heinz Gundlach, Ramamoorthy Ramesh, Valanoor Nagarajan, and Kristoff Szot for helpful and interesting discussions. We thank Christoph Buchal for critically reading the manuscript.

---